\documentclass[11pt,a4paper]{article}

\pdfoutput=1
\usepackage{jheppub}
\usepackage[latin1]{inputenc}
\usepackage{graphicx}
\usepackage{epsfig}
\usepackage{subfigure}
\usepackage{color}
\usepackage{comment}

\def\beq{\begin{equation}}
\def\eeq{\end{equation}}
\def\baq{\begin{eqnarray}}
\def\eaq{\end{eqnarray}}

\newcommand{\be}{\begin{equation}} 
\newcommand{\ee}{\end{equation}}
\newcommand{\bea}{\begin{eqnarray}} 
\newcommand{\eea}{\end{eqnarray}}
\newcommand{\nn}{\nonumber}
\newcommand{\bmp}{\noindent\begin{minipage}{16cm}}
\newcommand{\emp}{\end{minipage}\vskip 7mm} 
\def\lsim{\mathrel{\raise.3ex\hbox{$<$\kern-.75em\lower1ex\hbox{$\sim$}}}}
\def\gsim{\mathrel{\raise.3ex\hbox{$>$\kern-.75em\lower1ex\hbox{$\sim$}}}}

\newcommand{\intron}[1]{}

\subheader{\small{Preprint: HIP-2016-23/TH}}

\title{Feebly Interacting Dark Matter Particle as the Inflaton}

\author{Tommi Tenkanen}

\affiliation{University of Helsinki and Helsinki Institute of Physics, \\
                      P.O.~Box 64, FI-00014, Helsinki, Finland}
\emailAdd{tommi.tenkanen@helsinki.fi}

\abstract{We present a scenario where a $Z_2$-symmetric scalar field $\phi$ first drives cosmic inflation, then reheats the Universe but remains out-of-equilibrium itself, and finally comprises the observed dark matter abundance, produced by particle decays \`{a} la freeze-in mechanism. We work model-independently without specifying the interactions of the scalar field besides its self-interaction coupling, $\lambda\phi^4$, non-minimal coupling to gravity, $\xi\phi^2R$, and coupling to another scalar field, $g\phi^2\sigma^2$. We find the scalar field $\phi$ serves both as the inflaton and a dark matter candidate if $10^{-9}\lesssim \lambda\lesssim g\lesssim 10^{-7}$ and $3 \rm{keV} \lesssim m_{\rm \phi}\lesssim 85 \rm{MeV}$ for $\xi=\mathcal{O}(1)$. Such a small value of the non-minimal coupling is also found to be of the right magnitude to produce the observed curvature perturbation amplitude within the scenario. We also discuss how the model may be distinguished from other inflationary models of the same type by the next generation CMB satellites.

}


\begin{document}
\maketitle

%
\section{Introduction}

Extensions of the Standard Model of particle physics (SM) typically contain many scalar fields. Their role in explaining the observed curvature power spectrum and dark matter (DM) abundance, different early Universe phase transitions, matter-antimatter asymmetry, and many other phenomena have been studied extensively in the literature, as discussed, for example, in the recent reviews \cite{Mazumdar:2010sa,Morrissey:2012db,Martin:2013tda,Klasen:2015uma}. In this work, we study a class of beyond the SM scalar fields to address two major issues in cosmology: inflation and dark matter.

During the years 2009--13, the European Space Agency's Planck satellite measured properties of the Cosmic Microwave Background (CMB) and either supported, constrained, or even ruled out many scenarios of the early Universe physics. In particular, the Planck results -- together with many different astrophysical observations at different scales --- have shown overwhelming evidence for the existence of an unknown non-baryonic dark matter component, whose abundance in the Universe is now known to be $\Omega_{\rm DM} h^2\simeq 0.12$ \cite{Ade:2015xua}. How this abundance was produced in the early Universe is however still unknown, as no conclusive dark matter signals have shown up in experiments \cite{Klasen:2015uma}.

The Planck satellite placed bounds also on many inflationary scenarios by measuring the spectral index of primordial power spectrum to a high accuracy, $\Delta n_s= 0.0060$, and bounding the tensor-to-scalar ratio to $r<0.11$ \cite{Ade:2015lrj}. Among inflationary models the best fit to the Planck data is provided by different Starobinsky-like models, such as Higgs inflation \cite{Bezrukov:2007ep} or $s$-inflation \cite{Lerner:2009xg,Kahlhoefer:2015jma,Aravind:2015xst}, where a non-minimal coupling between gravity and quantum fields typically plays a crucial role.

In this work, we connect a Starobinsky-like inflationary model to dark matter production which occurs at a later stage in the history of the Universe. We consider a scenario where a $Z_2$-symmetric scalar field first drives cosmic inflation, then reheats the Universe but remains out-of-equilibrium itself, and finally comprises the observed dark matter abundance, produced by particle decays \`{a} la freeze-in mechanism \cite{McDonald:2001vt,Hall:2009bx,Yaguna:2011qn,Blennow:2013jba,Dev:2013yza,Elahi:2014fsa, Dev:2014tla,Kang:2015aqa,Nurmi:2015ema,Kainulainen:2016vzv}. As the $Z_2$ symmetric scalar field serves as both the inflaton and a FIMP ('Feebly Interacting Massive Particle') dark matter candidate, we name our scenario as the 'fimplaton' model\footnote{In contrast to our scenario, the standard $s$-inflation model \cite{Lerner:2009xg,Kahlhoefer:2015jma} and several other scenarios \cite{Bastero-Gil:2015lga} could be referred to as 'wimplaton' models, a term coined by the latter reference.}.

The paper is organized as follows: in Section \ref{model} we present the model and discuss general aspects of the phenomenology and requirements for the fimplaton scenario. Then, we present the scenario in a chronological order as it may have occured in the history of the Universe: first, in Sections \ref{inflationdynamics} and \ref{inflationobservables}, we study how the fimplaton with a non-minimal coupling to gravity drives inflation, then in Section \ref{reheating} we present a mechanism for reheating the Universe, and in Section \ref{freezein} discuss how this same scalar field comprises the observed DM abundance, produced by decays of other fields. Finally, in Section \ref{conclusions}, we conclude and present an outlook.

\section{The Model}
\label{model}

The model is specified by the potential

\be
\label{potential}
V(\phi,\sigma) = \frac{\mu_{\rm \sigma}^2}{2}\sigma^{\dagger}\sigma+\frac{\mu_{\rm \phi}^2}{2}\phi^2+\frac{\lambda_{\rm \sigma}}{4}(\sigma^{\dagger}\sigma)^2+\frac{\lambda_{\rm \phi}}{4}\phi^4+\frac{g}{2}\phi^2 \sigma^{\dagger}\sigma + V_{\rm gravity} ,
\ee
where both $\phi$ and $\sigma$ are scalar particles. We assume $\phi$ to be a real singlet but allow $\sigma$ to be charged under the Standard Model symmetries\footnote{It would be particularly interesting to study whether the scenario could be realized within a Higgs portal model where $\sigma$ is the SM Higgs and $\phi$ a singlet scalar. As our purpose is to present a new scenario, we do not restrict ourselves to this particular model.}. The term $V_{\rm gravity}$, including non-minimal couplings to gravity, is left unspecified until Section \ref{inflation}.

In the following, we take $\sigma$ to be a scalar which couples very weakly to $\phi$, $g\ll 1$, but sufficiently strongly to the SM particles, so that it becomes part of the SM heat bath during the Hot Big Bang era, while $\phi$ does not. The scalar $\phi$ we take to be the field responsible for driving inflation and later comprising the observed DM abundance. Stability of the DM particle is ensured by a $Z_2$-symmetry of the scalar potential. We also assume the physical masses satisfy $m_{\rm \sigma}>2m_{\rm \phi}$, so that the $\sigma$ field can decay to $\phi$ particles and produce the observed DM abundance by the freeze-in mechanism. Despite this mass hierarchy, we will show it is possible to produce a large amount of $\sigma$ particles out from a $\phi$ condensate during the reheating era.

The key requirement for the freeze-in production of DM is to assume that the DM particles had not become in thermal equilibrium with other particles before  production of the observed DM abundance at $T\simeq m_{\rm \sigma}$ \cite{McDonald:2001vt,Hall:2009bx}. Assuming thermal equilibrium within the visible sector, writing the Friedmann equation as

\be
H = \sqrt{\frac{\pi^2g_*}{90}}\frac{T^2}{M_{\rm P}} ,
\ee
where $M_{\rm P}$ is the reduced Planck mass, estimating $\langle\sigma_{\sigma\sigma\rightarrow \phi\phi} v\rangle\simeq g^2/T^2$, and using the usual expression for the number density of relativistic particle species, $n_{\rm \sigma}=\zeta(3)T^3/\pi^2$, we find that no thermalization of $\phi$ with other particle species occurs before $T\simeq m_{\rm \sigma}$ if 

\be
\label{thermalizationcond}
g \lesssim \sqrt{\frac{24.8\sqrt{g_*(m_{\rm \sigma})}m_{\rm \sigma}}{M_{\rm P}}} \simeq 10^{-7}\left(\frac{m_{\rm \sigma}}{125 \rm{GeV}}\right)^{1/2} ,
\ee
where the latter expression applies for $g_*(m_{\rm \sigma})\simeq 100$, and all quantities are evaluated at $T=m_{\rm \sigma}$. As a conservative benchmark value, we take $g\lesssim 10^{-7}$ in the following considerations. 

We further assume that also the $\phi$ self-interaction coupling takes a very small value, $\lambda_{\rm \phi}\lesssim g$, in line with the very small portal coupling. In fact, the chosen hierarchy of couplings is a necessary condition to ensure that the $\phi$ field indeed reheats the SM sector after inflation instead of decaying into its own quanta and remaining out-of-equilibrium forever, as we will show in Section \ref{reheating}. Along the lines of Ref. \cite{DeSimone:2008ei} and in line with the other small couplings, we also assume the higher-dimensional operators of the form

\be
\mathcal{L} = \mathcal{L}_4 + \lambda \phi^4\sum_{n>0}a_n\left(\frac{\phi}{\Lambda}\right)^n + \xi\phi^2R\sum_{n>0}b_n\left(\frac{\phi}{\Lambda}\right)^n + \dots ,
\ee
are suppressed, $a_n, b_n \ll 1$, and play no role in inflationary dynamics. Here $\mathcal{L}_4$ is the dimension four effective Lagrangian, $\Lambda\sim M_{\rm P}$, and $R$ is the Ricci scalar.

Finally, we neglect the renormalization group (RG) running of couplings and simply assume that at least the $\phi$-direction of the potential \eqref{potential} remains stable up to the Planck scale. While the RG running can in some parts of the parameter space render the inflaton potential unstable or affect predictions for inflationary observables, such as the spectral index $n_s$ or the tensor-to-scalar ratio $r$ \cite{Kahlhoefer:2015jma, Tenkanen:2016idg}, we leave a more detailed study of these aspects for concrete model setups, and in this work concentrate on predictions at the classical level only. 

\section{Cosmic inflation}
\label{inflation}

We begin by considering inflationary dynamics. Following closely the notation and prescription of Ref.'s \cite{Kahlhoefer:2015jma,Tenkanen:2016idg}, we present the part of the Lagrangian which couples non-minimally to gravity, and then discuss general inflationary dynamics and observables in Sections \ref{inflationdynamics} and \ref{inflationobservables}. In Section \ref{reheating}, we present a mechanism for reheating the Universe without allowing the fimplaton to become in thermal equilibrium with the SM fields.

\subsection{Inflationary dynamics}
\label{inflationdynamics}

We take $V_{\rm gravity}=\frac12(\xi_{\rm \phi}\phi^2 + \xi_{\rm \sigma}\sigma^2)R$, so that the Jordan frame action is

\be
\label{nonminimal_action}
S_J = \int d^4x \sqrt{-g}\bigg(\frac12\partial_{\mu}\sigma\partial^{\mu}\sigma + \frac12\partial_{\mu}\phi\partial^{\mu}\phi - \frac{1}{2}M_{\rm P}^2R-\frac12\xi_{\rm \sigma}\sigma^2R -\frac12\xi_{\rm \phi}\phi^2R - V(\phi,\sigma) \bigg),
\ee
where $V(\phi,\sigma)$ is the scalar potential \eqref{potential}. While other types of gravitational couplings, such as $\alpha R^2$, can also play important role in inflationary dynamics (see e.g. \cite{Salvio:2015kka,Calmet:2016fsr}), restricting to this simple non-minimal coupling between the Ricci scalar $R$ and scalar fields is motivated by the analysis of quantum corrections in a curved background which have been shown to generate such terms even if $\xi_i$ are initially set to zero \cite{Birrell:1982ix}.

The non-minimal couplings appearing in \eqref{nonminimal_action} can be removed by the usual conformal transformation, $\tilde{g}_{\mu\nu} = \Omega^2 g_{\mu\nu}$, where
\be
\label{Omega}
\Omega^2\equiv 1+\frac{\xi_{\rm \phi} \phi^2}{M_{\rm P}^2} + \frac{\xi_{\rm \sigma} \sigma^2}{M_{\rm P}^2}.
\ee
By then performing a field redefinition,
\be
\label{h_chi}
\frac{d\chi_{\rm \beta}}{d\beta} = \sqrt{\frac{\Omega^2+6\xi^2_{\rm \beta}\beta^2/M_{\rm P}^2}{\Omega^4}},
\ee
where $\beta=\phi,\sigma$, we obtain the so-called Einstein frame action
\be
\begin{aligned}
S_E =& \int d^4x \sqrt{-\tilde{g}}\bigg(-\frac{1}{2}M_{\rm P}^2\tilde{R} + \frac{1}{2}{\tilde{\partial}}_{\mu}\chi_{\rm \sigma}{\tilde{\partial}}^{\mu}\chi_{\rm \sigma} + \frac{1}{2}{\tilde{\partial}}_{\mu}\chi_{\rm \phi}
{\tilde{\partial}}^{\mu}\chi_{\rm \phi} \\
&+ A(\chi_{\rm \phi}, \chi_{\rm \sigma}){\tilde{\partial}}_{\mu}\chi_{\rm \sigma}{\tilde{\partial}}^{\mu}\chi_{\rm \phi} - U(\chi_{\rm \phi},\chi_{\rm \sigma})  \bigg),
\label{EframeS}
\end{aligned}
\ee
where 

\be
\label{Upotential}
U(\chi_{\rm \phi},\chi_{\rm \sigma}) = \Omega^{-4}V(\phi(\chi_{\rm \phi}),\sigma(\chi_{\rm \sigma})) ,
\ee 
and

\be
A(\chi_{\rm \phi}, \chi_{\rm \sigma}) = \frac{6\xi_{\rm \sigma}\xi_{\rm \phi}}{M_{\rm P}^2\Omega^4}\frac{d\phi}{d\chi_{\rm \phi}}\frac{d\sigma}{d\chi_{\rm \sigma}}\phi\sigma .
\ee

In the following, we will consider the scenario where inflation occurs in the $\phi$-direction. The scenario is similar to the so-called $s$-inflation \cite{Lerner:2009xg}. Consistency of this scenario requires that the minimum of the potential at large $\phi$ and $\sigma$ is very close to the $\sigma=0$ direction. This is true if $\lambda_{\rm \phi}/\xi^2_{\rm \phi}\ll \lambda_{\rm \sigma}/\xi^2_{\rm \sigma}$, which is easily satisfied for the values we will discuss below, $\lambda_{\rm \phi}\ll 1, \xi_{\rm \phi}\simeq 1$, and for $\lambda_{\rm \sigma}, \xi_{\rm \sigma}$ not too different from each other. In that case, the kinetic terms of the scalar fields are canonical as $A(\chi_{\rm \phi}, \chi_{\rm \sigma})=0$, and the analysis of inflationary dynamics can be performed in the usual way.

By taking into account only the highest order terms in the Jordan frame potential, $V(\phi,\sigma)=\lambda_{\rm \phi} \phi^4/4$, the Einstein frame potential becomes at large field values
\be
\label{chipotential}
U(\chi_{\rm \phi}) \simeq \frac{\lambda_{\rm \phi} M_P^4}{4\xi_{\rm \phi}^2}\left(1+\exp\left(-\frac{2\sqrt{\xi_{\rm \phi}}\chi_{\rm \phi}}{\sqrt{6\xi_{\rm \phi}+1}M_P} \right) \right)^{-2} ,
\ee
which is a sufficiently flat potential to support inflation at $\chi_{\rm \phi}\gg M_{\rm P}$, or equivalently at 
$\phi\gg M_{\rm P}/ \xi_{\rm \phi}^{1/2}$. Note that for $s$-inflation-type models the scale of perturbative unitarity breaking is always higher than the field value during inflation, provided that $\xi_{\rm \sigma}$ is small compared to $\xi_{\rm \phi}$ \cite{Kahlhoefer:2015jma}. In our scenario this requires $\xi_{\sigma} < 1$, and as the value of $\xi_{\sigma}$ can be chosen freely, we assume this to be always the case.

\subsection{Inflationary observables}
\label{inflationobservables}

The inflationary dynamics is characterized by the usual slow-roll parameters, which are defined in terms of the Einstein frame potential by

\bea
\label{epsiloneta}
\epsilon &\equiv& \frac{1}{2}M_P^2 \left(\frac{1}{U}\frac{{\rm d}U}{{\rm d}\chi_{\rm \phi}}\right)^2 \simeq \frac{8M_{\rm P}^4}{(6\xi_{\rm \phi}^2+\xi_{\rm \phi})\phi^4} ,\\ \nn
\eta &\equiv& M_P^2 \frac{1}{U}\frac{{\rm d}^2U}{{\rm d}\chi_{\rm \phi}^2} \simeq -\frac{8M_{\rm P}^2}{(6\xi_{\rm \phi}+1)\phi^2} ,
\eea
where the approximate values hold for $\phi \gg M_{\rm P}/ (6\xi_{\rm \phi}^2+\xi_{\rm \phi})^{1/2}$. In the following we solve for inflationary observables numerically by using the more accurate results provided by \eqref{h_chi} and \eqref{Upotential} but show also the approximative values to illustrate the parametric dependence of results.

\begin{figure}
\begin{center}
\includegraphics[width=.465\textwidth]{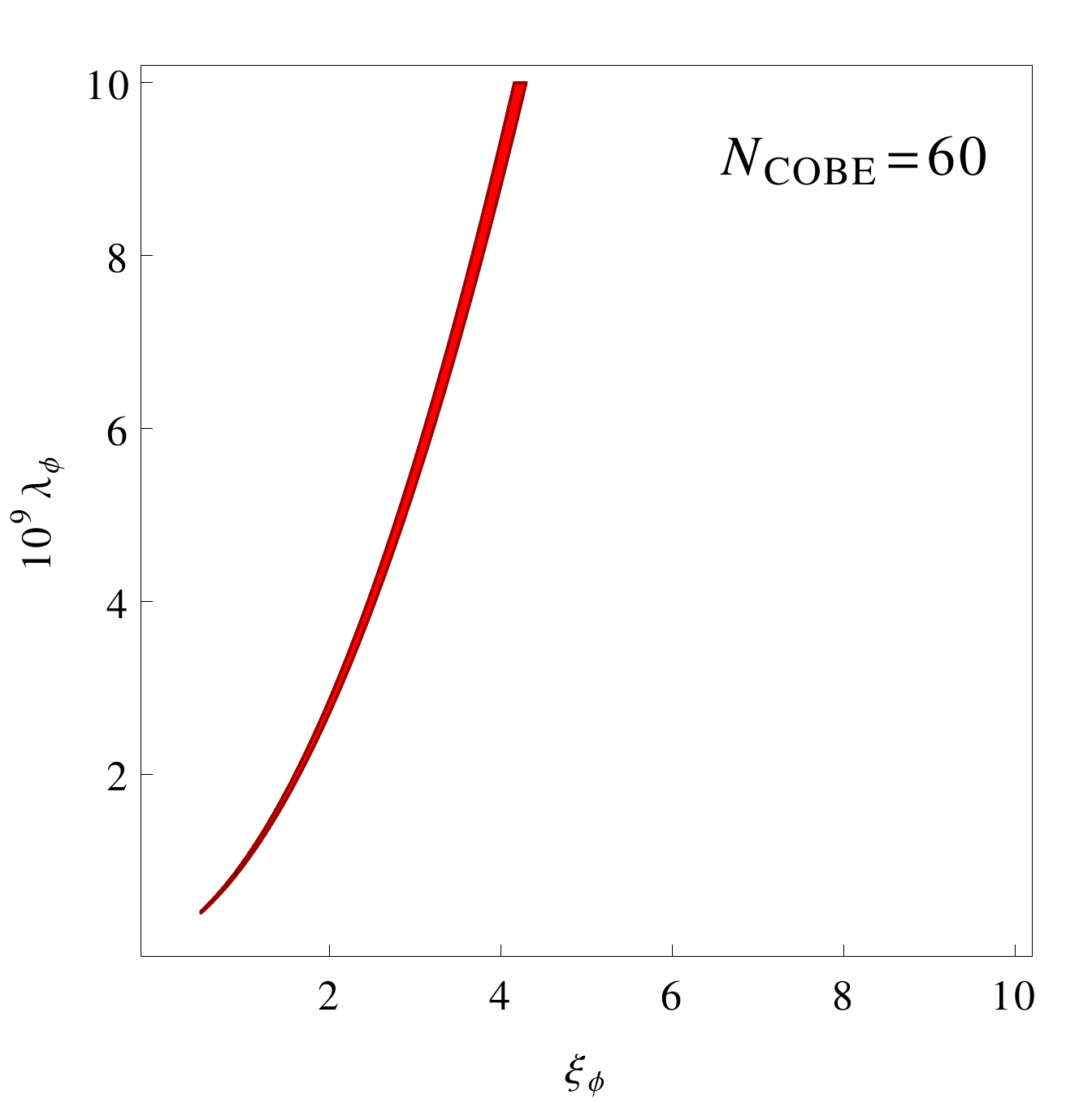}
\includegraphics[width=.45\textwidth]{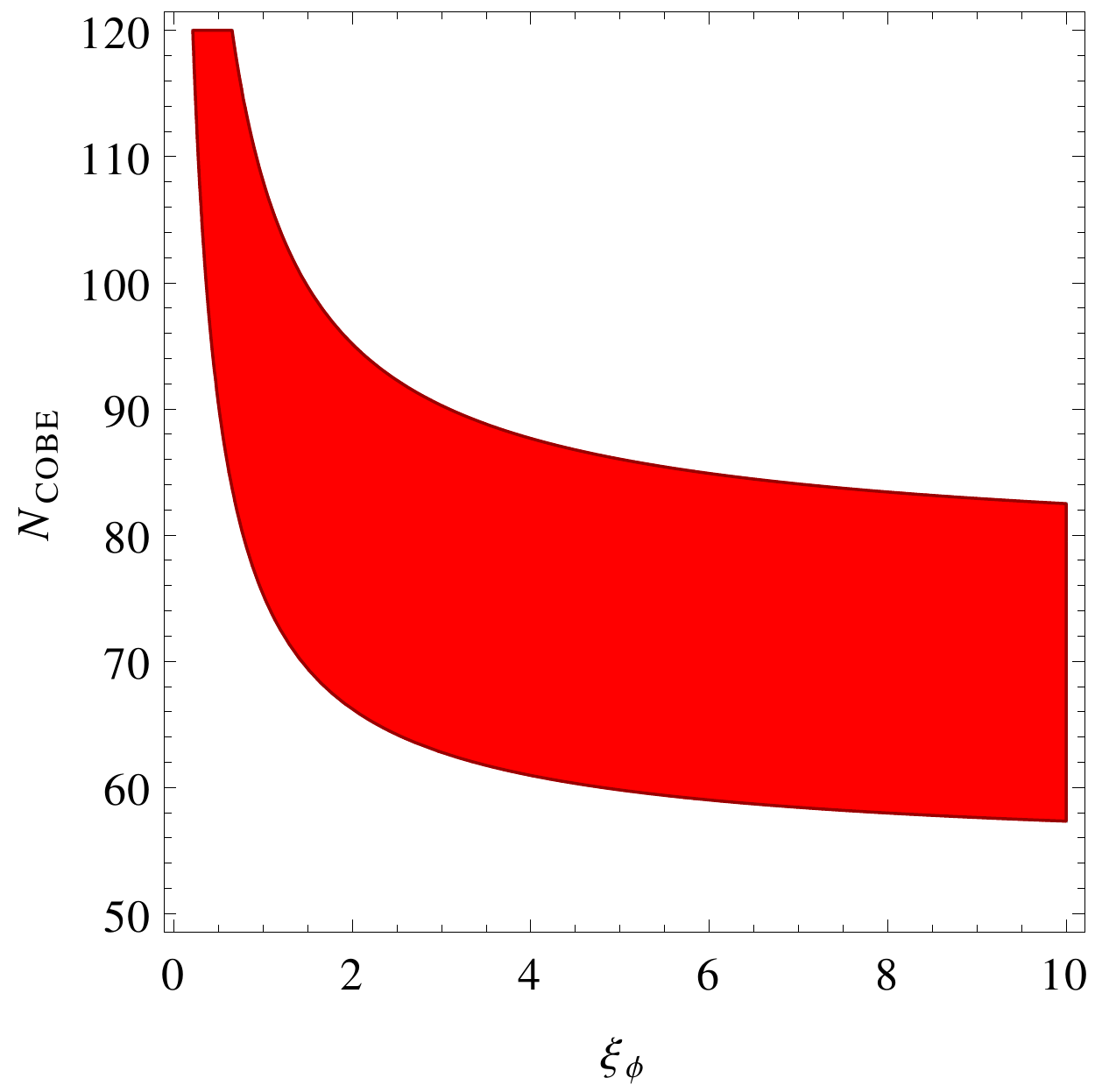}
\caption{Left panel: The required value of the scalar self-interaction coupling $\lambda_{\rm \phi}$ to produce the observed curvature perturbation amplitude, $\mathcal{P}_{\mathcal{R}}=(2.139\pm 0.063)\times 10^{-9}$, as a function of the non-minimal gravity coupling $\xi_{\rm \phi}$. Right panel: Inflationary observables in the $(\xi_{\rm \phi}, N_{\rm COBE})$--plane. In the red region the model satisfies both the Planck $1\sigma$ bound on spectral index, $n_s=0.9677\pm 0.0060$, and the Planck $2\sigma$ bound on tensor-to-scalar ratio, $r<0.11$.}
\end{center}
\label{inflationplot}
\end{figure}

The slow-roll inflation ends when $\epsilon\simeq 1$, giving

\be
\label{phiend}
\phi^2_{\rm end} \simeq \sqrt{\frac{8}{6\xi_{\rm \phi}^2+\xi_{\rm \phi}}}M_{\rm P}^2 ,
\ee
for the field value at the end of inflation. This allows us to calculate the number of inflationary e-folds, $N\equiv \ln(a_{\rm end}/a)$, for the change of inflationary field values from some initial $\phi_0$ to $\phi_{\rm end}$,

\be
\label{e-folds}
N = \frac{1}{M_{\rm P}^2}\int_{\phi_{\rm end}}^{\phi_0}U\left(\frac{dU}{d\phi}\right)^{-1}\left(\frac{d\chi_{\rm \phi}}{d\phi}\right)^2 d\phi \simeq \frac{6\xi_{\rm \phi}+1}{8M_{\rm P}^2}\left(\phi_0^2 - \phi^2_{\rm end} \right) .
\ee
The COBE normalization requires \cite{Lyth:1998xn}

\be
\label{cobe}
\frac{U}{\epsilon} \simeq \frac{M_{\rm P}^4\lambda_{\rm \phi}}{4\xi_{\rm \phi}^2}\frac{(6\xi_{\rm \phi}^2+\xi_{\rm \phi})\phi_{\rm COBE}^4}{8M_{\rm P}^4} = (0.0267\pm 0.0002)^4M_{\rm P}^4 ,
\ee
to obtain the measured amplitude of curvature power spectrum, $\mathcal{P}_{\mathcal{R}}=(2.139\pm 0.063)\times 10^{-9}$ (68\% confidence level) \cite{Ade:2015lrj}. Here $\phi_{\rm COBE}$ is the field value at the time there was $N_{\rm COBE}$ e-folds left of inflation. Solving for $N_{\rm COBE}$ from \eqref{e-folds}, we can express the requirement \eqref{cobe} in terms of e-folds,

\be
\label{xi}
\frac{2\lambda_{\rm \phi} N_{\rm COBE}^2}{6\xi_{\rm \phi}^2+\xi_{\rm \phi}} \simeq (0.0267\pm 0.0002)^4.
\ee
It is the equation \eqref{xi} that determines the required value of the non-minimal coupling $\xi_{\rm \phi}$ in terms of $\lambda_{\rm \phi}$ and $N_{\rm COBE}$. Because we assume $\lambda_{\rm \phi}\lesssim g\lesssim 10^{-7}$ to reheat the Universe without thermalizing the fimplaton with the SM bath (see Section \ref{reheating}), we find $\xi_{\rm \phi}\lesssim 10$ for the allowed values of the non-minimal coupling, as depicted in the left panel of Figure \ref{inflationplot}. In the following we take $\xi_{\rm \phi}>1$ to simplify the reheating analysis. This choice corresponds to $\lambda_{\rm \phi}\gtrsim 10^{-9}$.

As was shown in \cite{Tenkanen:2016idg}, the number of e-folds in a feebly coupled $s$-inflation-type model is given by
\be
\label{efolds}
N \simeq 60-\frac{1}{12}\ln\left(\lambda_{\rm \phi}N^4 \right),
\ee
which for $10^{-9}\lesssim \lambda_{\rm \phi}\lesssim 10^{-7}$ gives $N\simeq 60$. For the spectral index, $n_s(k)-1\equiv d\mathcal{P}_{\mathcal{R}}/d\ln k \simeq -6\epsilon+2\eta$, and tensor-to-scalar ratio, $r\equiv \mathcal{P}_{\mathcal{T}}/\mathcal{P}_{\mathcal{R}} \simeq 16\epsilon$, we then obtain the following numerical results

\bea
\label{nsr}
n_s &\simeq& 0.962 , \nn \\
r &\simeq& 0.0042,
\eea
where the slow-roll parameters have been evaluated at $N=60$, and at the minimum value of the corresponding non-minimal coupling, $\xi_{\rm \phi}\simeq 5$ (see the right panel of Figure \ref{inflationplot}). To see how predictions vary for different parameter values, we solve the inflationary observables numerically for $59<N<61$ and $4<\xi_{\rm \phi}<10$ to obtain

\be
0.960 < n_s < 0.964, \hspace{1cm} 0.0036 < r < 0.0045 .
\ee

The numerical results for the spectral index $n_s$ and tensor-to-scalar ratio $r$  are not only compatible with the Planck results $n_s=0.9677\pm 0.0060$ ($68\%$ confidence level), $r<0.11$ ($95\%$ confidence level) \cite{Ade:2015lrj} but differ from other inflationary models of the same type. For example, for Higgs inflation \cite{Lerner:2011ge} 

\be
0.964 < n_s < 0.965, \hspace{1cm} 0.0033 < r < 0.0037, 
\ee
and for $s$-inflation \cite{Lerner:2011ge}

\be
0.964 < n_s < 0.966,\hspace{1cm} 0.0032 < r < 0.0036 .
\ee

Despite the fact that the predicted value of tensor-to-scalar ratio is almost two orders of magnitude lower than the current upper bound, there is hope in detecting it with the next generation CMB satellites which plan to measure $r$ to an accuracy $\Delta r=10^{-3}$, such as PIXIE \cite{Kogut:2011xw} and LiteBIRD \cite{Matsumura:2013aja}. It will be interesting to see whether they will also be able to distinguish between different models of the same type.

\subsection{Reheating after inflation}
\label{reheating}

Reheating in $s$-inflation-type models has been discussed extensively in Ref.'s \cite{Lerner:2011ge, Tenkanen:2016idg} (see also \cite{Bezrukov:2008ut}). In this section, we review the main results of reheating in $s$-inflation for $\xi_{\rm \phi}>1$ and discuss how the fimplaton $\phi$ can reheat the Universe but remain out-of-equilibrium itself. We need to ensure that reheating occurs at $T_{\rm RH}> m_{\rm \sigma}$, or otherwise the computation becomes inconsistent with the assumption that $\phi$ is a frozen-in dark matter particle produced by $\sigma\rightarrow \phi\phi$ decays at $T\simeq m_{\rm \sigma}$, as we will discuss in Section \ref{freezein}.

After inflation the Einstein frame $\chi_{\rm \phi}$ condensate begins to oscillate with an initial field value $\chi_*(\phi_{\rm end})$, where $\phi_{\rm end}$ is determined by Eq. \eqref{phiend}. The field oscillates first in a quadratic potential with $\chi \propto a^{-3/2}$, until a transition into quartic potential occurs at $\phi\simeq \sqrt{2/3}M_{\rm P}/\xi_{\rm \phi}$ \cite{Bezrukov:2008ut}. After this $\Omega^2\rightarrow 1$, and the Einstein and Jordan frames become equivalent. In the following we will use the Jordan frame notation.

After transition the homogeneous fimplaton condensate evolves as

\be 
\phi_0(t)=\Phi_0(t){\rm cn}(0.85\lambda_{\rm \phi}^{1/2}\Phi_0(t) t, 1/\sqrt{2}) ,
\ee
where cn is the Jacobi cosine, $\Phi_0$ a time-dependent oscillation amplitude, and $t$ the cosmic time. As shown for $s$-inflation in \cite{Tenkanen:2016idg}, it is the quartic regime where reheating occurs if the couplings between the SM sector and inflaton are very weak, and we have indeed checked that there is no significant particle production in the quadratic regime for $g \ll 10^{-4}$ if $\lambda_{\rm \phi}\geq 10^{-9}$.

The oscillating background generates an additional mass term for $\phi$ and $\sigma$ particles
\be
\label{adiabatic_masses}
\begin{aligned}
M_{\rm \phi}^2 &= \mu^2_{\rm \phi} + 3\lambda_{\rm \phi} \phi_0(t)^2, \\
M_{\rm \sigma}^2 &= \mu_{\rm \sigma}^2 + \frac{g}{2} \phi_0(t)^2 , \\
\end{aligned}
\ee
where we have assumed that no thermal corrections arise whose contribution to mass terms, $\mu_i=\mu_i(T)$, could block the decay of the fimplaton condensate\footnote{If $\sigma$ is the SM Higgs and $\phi$ a portal scalar, special care should be taken in analyzing formation of thermal corrections.}. The decay rates of the condensate energy density induced by the interactions $\lambda_{\rm \phi} \phi_0(t)^2 \phi^2$ and $g \phi_0(t)^2 \sigma^2$ are given by \cite{Nurmi:2015ema,Kainulainen:2016vzv}

\be
\label{decayrates}
\begin{aligned}
\Gamma_{\phi_0\rightarrow \phi\phi} &= \frac{9\lambda_{\rm \phi}^2 \omega}{8\pi\rho_{\rm \phi_0}}\sum_{n=1}^{\infty}n|\zeta_n |^2\sqrt{1-\left(\frac{M_{\rm \phi}}{n\omega}\right)^2} ,  \\
\Gamma_{\phi_0\rightarrow \sigma\sigma} &=\frac{g^{2}\omega}{8\pi\rho_{\rm \phi_0}}\sum_{n=1}^{\infty}n|\zeta_n |^2 \sqrt{1-\left(\frac{M_{\rm \sigma}}{n\omega}\right)^2} ,
\end{aligned}
\ee
where 

\be
\label{s_fourier}
\phi_0(t)^2 = \sum_{n=-\infty}^{\infty}\zeta_n e^{-i2\omega nt} ,
\ee
and $\omega\simeq 0.85\lambda_{\rm \phi}^{1/2}\Phi_0$ is the oscillation frequency of $\phi_0$. Here $\rho_{\rm \phi_0}=\lambda_{\rm \phi}\Phi_0^4/4$ is the energy density of $\phi_0$. Finally, we average the decay rates over one oscillation cycle. We note that the semi-perturbative decay rates \eqref{decayrates} provide not only a useful calculation method but also account for adiabatic mass terms \eqref{adiabatic_masses} and are thus expected to describe dynamics of reheating to a sufficient accuracy.

Because we have assumed that the two sectors never become in thermal equilibrium with each other, requiring $g\lesssim 10^{-7}$, the fimplaton condensate has to decay to $\sigma$ particles which further reheat the SM sector -- instead of decaying into $\phi$ particles which remain out of thermal equilibrium forever\footnote{As shown in \cite{Tenkanen:2016jic}, annihilations of thermally decoupled $Z_2$ symmetric scalars cannot heat up the SM sector in a way consistent with the Big Bang Nucleosynthesis.}. Therefore, we need

\be
\label{decayRequirement}
\Gamma_{\phi_0\rightarrow \sigma\sigma}>\Gamma_{\phi_0\rightarrow \phi\phi} .
\ee

In the quartic regime the kinematic condition is $n^2\omega^2>M_{\rm \sigma}^2$, \eqref{decayrates}, which can be fulfilled with a sufficiently large $n$ even if $m_{\rm \sigma}>m_{\rm \phi}$. By neglecting the bare masses in the quartic regime, we find that the requirement \eqref{decayRequirement} is satisfied for $g\geq 3\lambda_{\rm \phi}$, and the dominant decay rate becomes

\be
\Gamma_{\phi_0\rightarrow \sigma\sigma} \simeq 0.002g^2\lambda_{\rm \phi}^{-1/2}\Phi_0 .
\ee

As we are interested in a scenario where $\phi$ is not only the inflaton but also a frozen-in dark matter candidate, we require the reheating temperature to satisfy $T_{\rm RH}\gtrsim m_{\rm \sigma}$. The fimplaton condensate decays at $\Gamma_{\phi_0\rightarrow \sigma\sigma}\simeq H$, giving 

\be
\Phi_0 = 0.007\frac{g^2}{\lambda_{\rm \phi}}M_{\rm P} ,
\ee 
for the field value at the time of condensate decay. Here we used

\be
H = \sqrt{\frac{\lambda_{\rm \phi}}{12}}\frac{\Phi_0^2}{M_{\rm P}} ,
\ee
which can be derived by using the fimplaton energy density, $\rho_{\rm \phi_0}=\lambda_{\rm \phi}\Phi_0^4/4$, in the Friedmann equation $3H^2M_{\rm P}^2=\rho$. Equating then the fimplaton energy density with the energy density of the heat bath, $\pi^2g_*T^4/30$, at the time of the fimplaton decay gives 

\be
\label{rhtemp}
T_{\rm RH} = 0.002\left(\frac{g_*(T_{\rm RH})}{10^2}\right)^{-1/4}g^2\lambda_{\rm s}^{-3/4}M_{\rm P} ,
\ee
where $g_*(T_{\rm RH})$ is the effective number of degrees of freedom in the heat bath at the time of reheating. For $g\geq3\lambda_{\rm \phi}, \lambda_{\rm \phi}\geq 10^{-9}$, $g_*\simeq100$, the reheating temperature \eqref{rhtemp} then sets a conservative upper limit on the $\sigma$ mass

\be
\label{sigmamass}
m_{\rm \sigma}< T_{\rm RH}\lesssim 3\times 10^5 {\rm GeV} , 
\ee
if the dark matter relic density is to be produced by $\sigma \rightarrow \phi\phi$ at $T\simeq m_{\rm \sigma}$. In Section \ref{freezein}, we discuss how this upper bound relates to a bound on dark matter mass and coupling values.

\section{Dark Matter production}
\label{freezein}

Finally, we turn to dark matter production. We have assumed the portal coupling takes a very small value, $g\lesssim 10^{-7}$, which prevents the fimplaton $\phi$ from becoming into thermal equilibrium with the bath particles. Thus, the DM relic density has to be produced by scalar decays\footnote{Also $2\leftrightarrow 2$ scatterings are known to contribute to the DM freeze-in yield. We neglect these processes for simplicity.} $\sigma\rightarrow \phi\phi$ at $T\simeq m_{\rm \sigma}$ instead of the standard thermal freeze-out mechanism where the DM number density freezes to a constant value when different annihilation processes, such as $\phi\phi \rightarrow \sigma\sigma$, can not compete with the expansion rate of the Universe any more.

The freeze-in production of dark matter has been studied extensively in e.g. \cite{McDonald:2001vt,Hall:2009bx,Yaguna:2011qn,Blennow:2013jba,Dev:2013yza,Elahi:2014fsa, Dev:2014tla,Kang:2015aqa,Nurmi:2015ema,Kainulainen:2016vzv}. The canonical result for the frozen-in DM abundance is \cite{Hall:2009bx}

\be
\label{freezeinabundance}
\Omega_{\rm \phi}h^2 = \frac{1.09\times10^{27}}{g_{*S}\sqrt{g_*}}\frac{m_{\rm \phi}\Gamma_{\sigma\rightarrow \phi\phi}}{m_{\rm \sigma}^2} ,
\ee
where $g_{*S}$ and $g_*$ are, respectively, the effective numbers of entropy and energy density degrees of freedom at the time the DM density freezes in. Taking $\Gamma_{\sigma\rightarrow \phi\phi}=g^2m_{\rm \sigma}/(8\pi)$ and $g_{*S}\simeq g_*$, the result \eqref{freezeinabundance} can be written as

\be
\label{freezeinabundance2}
\frac{\Omega_{\rm \phi}h^2}{0.12} = 3.6\times10^{23}g^2\left(\frac{10^2}{g_*}\right)^{3/2}\left(\frac{m_{\rm \phi}}{m_{\rm \sigma}}\right) .
\ee
Due to very feeble DM self-interactions in the fimplaton scenario, $\lambda_{\rm \phi}\simeq \mathcal{O}(10^{-9})$, the abundance produced by scalar decays, \eqref{freezeinabundance2}, is indeed the final DM abundance, and no thermal freeze-out operating in the dark matter sector --  as recently studied in e.g. \cite{Bernal:2015ova,Bernal:2015xba,Heikinheimo:2016yds} -- need to be considered.

Taking $\phi$ to constitute all of the observed DM abundance gives us

\be
\label{massbound}
\frac{m_{\rm \phi}}{m_{\rm \sigma}} \simeq 3\times 10^{-24}g^{-2} \left(\frac{g_*}{10^2}\right)^{3/2} .
\ee
The result \eqref{massbound} allows for deriving bounds on the fimplaton mass. If the scalar $\sigma$ is not necessarily the SM Higgs but a heavy mediator between the SM and fimplaton sectors, we can use the upper bound on the $\sigma$ mass, \eqref{sigmamass}, to obtain

\be
m_{\rm \phi}\lesssim 6\times 10^{-27}\left(\frac{g_*(m_{\rm \sigma})}{10^2}\right)^{3/2}\left(\frac{g_*(T_{\rm RH})}{10^2}\right)^{-1/4}\frac{g^2(T_{\rm RH})}{g^2(m_{\rm \sigma})}\lambda^{-3/4}_{\rm \phi}(T_{\rm RH}) M_{\rm P} ,
\ee
which is a strict upper bound on the fimplaton mass. Here we have explicitly written the scales where the couplings should be evaluated. 

\begin{figure}
\begin{center}
\includegraphics[width=.45\textwidth]{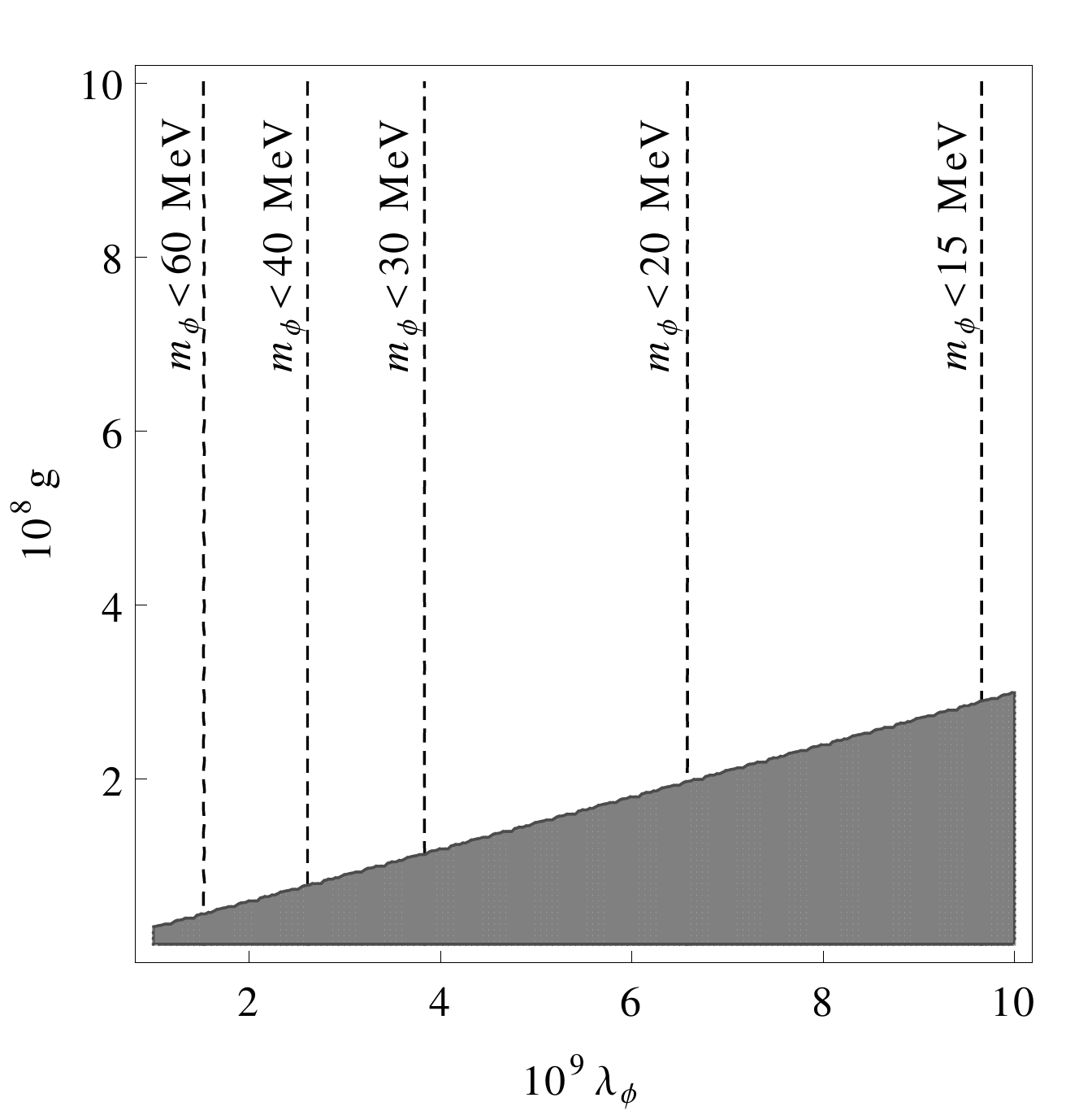}
\includegraphics[width=.455\textwidth]{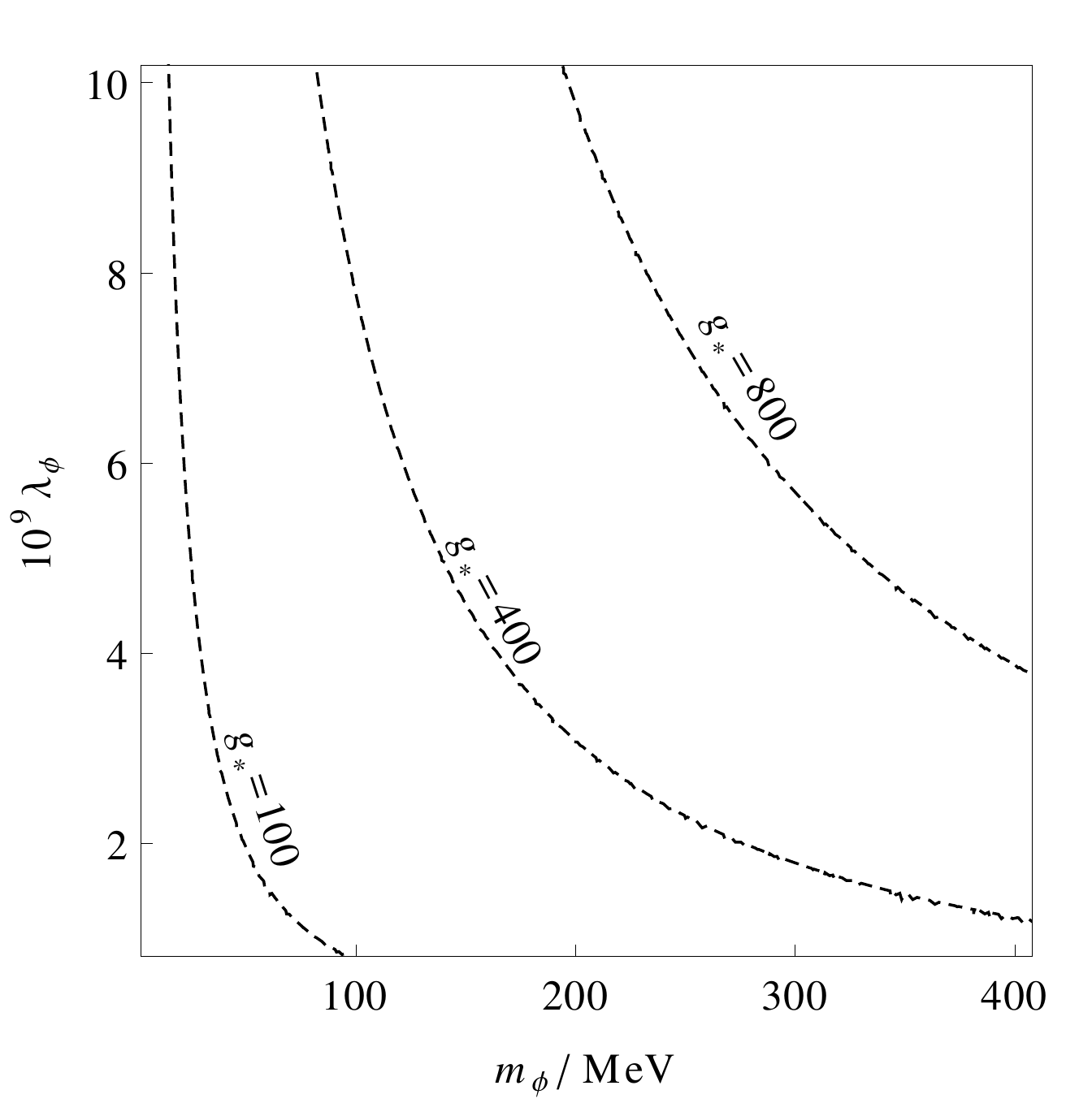}
\caption{Left panel: Bounds on the fimplaton mass from reheating and dark matter production in the $(\lambda_{\rm \phi}, g)$--plane for $g_*(T_{\rm RH})=g_*(m_{\rm \sigma})=100$. For a given value of the fimplaton self-interaction strength $\lambda_{\rm \phi}$, masses larger than the value in the corresponding contours are excluded. The gray region, $g\leq 3\lambda_{\rm \phi}$, is excluded by reheating dynamics. Right panel: The same mass bounds in the $(m_{\rm \phi},\lambda_{\rm \phi})$--plane for different values of $g_*(T_{\rm RH})=g_*(m_{\rm \sigma})$. For a given value of $g_*$, regions above the contours are excluded.}
\label{massboundfig}
\end{center}
\end{figure}

Neglecting the RG running, using $\lambda_{\rm \phi}\gtrsim 10^{-9}$, and assuming again that the total number of energy density and entropy degrees of freedom does not differ too much from $g_*\simeq 100$, we get $m_{\rm \phi}\lesssim 85$ MeV to be an absolute upper bound on the fimplaton mass. On the other hand, astrophysical observations of the Lyman-$\alpha$ forest impose a lower bound on dark matter mass, $m_{\rm DM}\gtrsim 3$ keV \cite{Viel:2013apy}, so that in total the fimplaton mass is bounded to 

\be
3 \rm{keV} \lesssim m_{\rm \phi}\lesssim 85 \rm{MeV} .
\ee
For $\xi\lesssim 1$ smaller values of $\lambda_{\rm \phi}>10^{-9}$, and therefore larger values of $m_{\rm \phi}$, become allowed but at the expense of having a large number of required e-folds, $N\gg 60$, see Figure \ref{inflationplot}. In this work we restrict to the more conventional value $N\simeq 60$.

The results are shown for different values of $\lambda_{\rm \phi}$ and $g_*$ in Figure \ref{massboundfig}. In particular, if $\sigma$ is the SM Higgs, the absolute upper bound on the fimplaton mass becomes $m_{\rm \phi}\lesssim 40$ keV irrespectively of $\lambda_{\rm \phi}$, see \eqref{massbound}. Here we used $g>10^{-9}$ and $g_*=106.75$ at the time of the DM freeze-in.

\section{Conclusions}
\label{conclusions}

In this work we have studied a scenario where a $Z_2$-symmetric scalar field, non-minimally coupled to gravity, drives cosmic inflation, reheats the Universe but remains \linebreak out-of-equilibrium itself, and later comprises the observed dark matter abundance, produced by particle decays \`{a} la freeze-in mechanism. As the $Z_2$-symmetric scalar field serves as both the inflaton and a FIMP ('Feebly Interacting Massive Particle') dark matter candidate, we have named our scenario as the 'fimplaton' model. 

Because we wanted to work as model-independently as possible, we did not specify the fimplaton's connection to the known Standard Model physics nor its interactions besides its self-interaction coupling, $\lambda_{\rm \phi}\phi^4$, non-minimal coupling to gravity, $\xi_{\rm \phi}\phi^2R$, and coupling to another scalar field, $g\phi^2\sigma^2$. It would be interesting to study whether already e.g. a simple Higgs portal model, where $\sigma$ is the SM Higgs 
and $\phi$ a portal scalar, could accommodate the fimplaton scenario.

We have shown the fimplaton model constitutes an interesting example of a scenario where even very small couplings can be responsible for both inflation, reheating, and production of the observed dark matter abundance. Although the (somewhat conservative) coupling and mass windows where the model works are relatively narrow,\linebreak $10^{-9}\lesssim \lambda_{\rm \phi}\lesssim g\lesssim 10^{-7}$, $3 \rm{keV} \lesssim m_{\rm \phi}\lesssim 85 \rm{MeV}$, the scenario is shown to provide a successful connection between cosmic inflation and dark matter abundance. Furthermore, as shown in Section \ref{inflationobservables}, the model may be distinguishable from other inflationary models of the same type, namely the Higgs inflation and $s$-inflation, by the next generation CMB satellites.

An interesting aspect of the fimplaton model is that to produce the observed curvature perturbation amplitude within the scenario, the non-minimal coupling has to take a relatively small value, $\xi_{\rm \phi}=\mathcal{O}(1)$. This is indeed a very small value, as Higgs and $s$-inflation models typically require $\xi=\mathcal{O}(10^4)$. As quantum corrections in a curved background have been shown to generate small non-minimal couplings even if they are initially set to zero, it would be particularly interesting to apply the fimplaton scenario to concrete model setups. As SM extensions typically contain many new scalar fields, studies of their role in both inflation and dark matter production together provide many new ways to extract information about SM extensions and physics of the early Universe in general.

\section*{Acknowledgements}
The author thanks V. Vaskonen for discussions and acknowledges financial support from the Research Foundation of the University of Helsinki.

\bibliography{fimpinflation.bib}

\end{document}